 \documentclass[prd,
 notitlepage,
nofootinbib,
amsfonts,
amssymb,
amsmath,
reprint,
 ]{revtex4-2}
\usepackage{color}
\usepackage{graphicx}
\usepackage{xcolor}
\usepackage{amsmath}
\usepackage{soul}
\definecolor{bluc}{cmyk}{1,0.3,0,0}
\definecolor{rossoCP3}{cmyk}{0,.88,.77,.40}
\definecolor{rosso}{cmyk}{0,1,1,0.4}
\definecolor{rossos}{cmyk}{0,1,1,0.55}
\definecolor{rossoc}{cmyk}{0,1,1,0.2}
\definecolor{verdes}{cmyk}{0,0.8,0,0}
\usepackage[normalem]{ulem}
\usepackage{hyperref}
\hypersetup{colorlinks, bookmarksopen, bookmarksnumbered, citecolor=verdes, linkcolor=bluc, pdfstartview=FitH, urlcolor=rossos}
\usepackage[utf8x]{inputenc}
\usepackage[T1]{fontenc}
\usepackage{subdepth}

\usepackage{multirow}
\usepackage[shortlabels]{enumitem}

\newcommand\rst{\bgroup\markoverwith{\textcolor{red}{\rule[0.5ex]{1pt}{1.4pt}}}\ULon}

\begin{document}
\title{Strong lensing by GUP-improved black holes}
\author{Javier Chagoya$^{1}$}
\email{javier.chagoya@fisica.uaz.edu.mx}
\author{I. D\'iaz-Saldaña$^{1}$}
\email{isaacdiaz@fisica.uaz.edu.mx}

\author{Benito Rodríguez$^{1}$}
\email{benito.rodriguez@fisica.uaz.edu.mx}

\author{Wilfredo Yunpanqui$^{1,2}$}
\email{w.yupanqui@fisica.uaz.edu.mx}
\affiliation{$^{1}$
 Unidad Acad\'emica de F\'isica, Universidad Aut\'onoma de Zacatecas, Calzada Solidaridad esquina con Paseo a la Bufa S/N 
 C.P. 98060, Zacatecas, M\'exico.
}
\affiliation{$^{2}$
 Departamento de F\'isica, Divisi\'on de Ciencias e Ingenier\'ias, Universidad de Guanajuato, Loma del Bosque 103, Le\'on, 37150, Guanajuato, M\'exico.
}
\date{\today}

\begin{abstract}
We employ Bozza’s method to calculate the deflection angle of light in 
presence 
of the strong gravitational field 
generated by an improved 
Schwarzschild-like black hole whose metric, regular throughout the entire spacetime, was derived using 
the improved Generalized Uncertainty Principle (GUP). This framework incorporates effective quantum gravity corrections that resolve the physical singularity inside the black hole, quantified by a model parameter $|\tilde{Q}_c|$. In addition, the event horizon, the photon sphere, and the shadow radius receive modifications characterized by a second model parameter $\tilde Q_b$. Using observational properties of the supermassive black holes Messier 87$^*$ and  Sagittarius A$^*$ reported by the Event Horizon Telescope, we derive constraints on the parameter $|\tilde{Q}_b|$, namely $0 \leq |\tilde{Q}_b| \leq 0.3$. To the best of our knowledge, these are the first constraints reported in the literature for this improved GUP parameter. Since  $|\tilde{Q}_c|$ does not play a significant role in the correction of the shadow radius, it was not possible to impose restrictions on its allowed values, however it is important to consider a non-zero $|\tilde Q_c|$ in order to avoid a black hole singularity.
\end{abstract}

\maketitle

\section{Introduction}
Quantum black holes constitute a fundamental arena for testing the interplay between general relativity and quantum mechanics, as they are expected to exhibit phenomena that lie beyond the scope of classical physics. As such, their study plays a pivotal role in the quest for a consistent theory of quantum gravity, offering a potential window into the underlying quantum nature of spacetime. This line of research is particularly timely given the growing likelihood that future astrophysical observations will detect signs of quantum gravity, particularly from black holes.

Several approaches have been developed to investigate black holes within the framework of quantum gravity. Among them, Loop Quantum Gravity (LQG)~\cite{thiemann2008modern} has provided fertile ground for the study of both the interior and the full spacetime structure of black holes. In this context, numerous analysis have focused on the canonical quantization of the black hole interior using Ashtekar-Barbero variables, where the dynamics are governed by an effective Hamiltonian subjected to a framework known as polymer quantization techniques~\cite{Abhay_Ashtekar_2003,PhysRevD.76.044016,PhysRevD.95.065026,PhysRevD.92.104029,FLORESGONZALEZ2013394,PhysRevLett.110.211301}. This procedure introduces a fundamental discreteness via a polymerization parameter, leading to a resolution of the classical singularity and, in the vacuum case, to a transition (or ``bounce'') from a black hole to a white hole~\cite{Corichi_2016,PhysRevD.78.064040,MORALESTECOTL2021168401,PhysRevD.76.104030,https://doi.org/10.1155/2008/459290,PhysRevD.103.084038,Sobrinho_2023}. Moreover, polymer quantization induces nontrivial quantum corrections to the classical constraint algebra, which can be interpreted as effective deformations at the semiclassical level. Alternative quantization schemes for black hole interiors have also been explored in the literature~\cite{Adeifeoba_2018,olmo2016classical,Husain_2005,PhysRevD.73.104009,jalalzadeh2012quantization,PhysRevD.57.2349,Almeida_2023,PhysRevD.60.024009,PhysRevLett.98.181301,PhysRevD.81.023528}.

The incorporation of gravitational effects into quantum measurement processes has motivated generalized formulations of the Heisenberg Uncertainty Principle (HUP). One of the most prominent outcomes of such considerations is the Generalized Uncertainty Principle (GUP), which introduces a minimal position uncertainty by modifying the standard uncertainty relation to account for high-energy (Planck-scale) deformations~\cite{PhysRevD.52.1108}. From a formal perspective, the GUP can also be interpreted as an alternative quantization scheme where the canonical commutation relations between position and momentum (or between generalized position and momentum, depending on the representation~\cite{PhysRevD.52.1108}) are deformed~\cite{PhysRevD.85.024016,PhysRevD.107.126009}, resulting in a minimal nonzero uncertainty in position and or momentum~\cite{BOSSO2018498}. Interestingly, recent results suggest that the GUP framework can emerge from generalized statistical mechanics; for instance, in~\cite{BIZET2023137636}, it is shown that GUP-type relations naturally arise when entropy is modeled via non-extensive forms that depend only on probability distributions.

Moreover, GUP relations have been independently derived through various theoretical considerations: in~\cite{G_Veneziano_1986}, string scattering at trans-Planckian energies is used to address quantum gravitational divergences; in~\cite{MAGGIORE199365}, a gedanken experiment is proposed for measuring the area of black hole apparent horizons within a quantum gravity setting; and in~\cite{SCARDIGLI199939}, the possibility of spacetime fluctuations at the Planck-scale is examined, leading to the notion that virtual micro-black holes may influence measurement processes. Despite the theoretical appeal of the GUP framework, several conceptual challenges and unresolved issues remain, as critically analyzed in \cite{Bosso_2023}, where the authors point out that these concerns are often overlooked or not adequately addressed in the literature.

An acceptable model of quantum gravity is expected to resolve, at least effectively, the singularities associated with black holes. In this context, in~\cite{Bosso_2021,MELCHOR2024116584} the authors investigate the interior dynamics of black holes from a semiclassical (or effective) perspective by introducing a deformed Poisson algebra inspired by GUP. This deformation involves parameters, commonly referred to as \emph{deformation parameters}, which modify the dynamical behavior inside the black hole. By choosing specific values for these parameters, one can explore how deformed algebra affects the dynamics of the system and physical observables.

The approach of studying aspects of classical dynamics through GUP-type deformations has been previously considered in the literature~\cite{Bosso_2021,PhysRevD.65.125028,PhysRevD.66.026003}, however, obtaining the classical limit of the corresponding commutation relations is more subtle than simply applying the replacement $\frac{[\hat{q}_i,\hat{p}_i]}{i\hbar} \to \{q_i, p_i\}$. In~\cite{CASADIO2020135558}, it was shown that it is not feasible to estimate any GUP-induced effect in the classical limit. In contrast~\cite{Bosso_2023}, consistent limits of the GUP framework are discussed, and it is argued that, contrary to the conclusions in~\cite{CASADIO2020135558}, it is indeed reasonable to consider corrections to classical dynamics arising from Planck-scale suppressed effects, although the interpretation of such corrections may be more nuanced than in the quantum regime. 

Recently, in~\cite{PhysRevD.111.024048}, the authors derived the full spacetime metric of an effective black hole inspired by GUP. They revisited a previous model for the black hole interior proposed in~\cite{Bosso}, and demonstrated that a naive extension of its metric to the entire spacetime leads to serious issues in the asymptotic region. To address this, they introduced an ``improved scheme'' inspired by LQG techniques, in which the deformation parameters are promoted to momentum-dependent functions.  Within this improved framework, the interior solution is consistently extended to the entire spacetime, producing a metric that is asymptotically flat and whose associated Kretschmann scalar remains everywhere regular. As in previous studies, this construction was carried out in an effective approach.

In recent years, the deflection of light due to strong gravitational fields has received increased attention as a tool for detecting observational signatures of modified gravity. The study of strong lensing near the Schwarzschild black hole dates back to 1959, when Darwin obtained an exact expression for the deflection angle~\cite{doi:10.1098/rspa.1959.0015}. 
  Recently, the Event Horizon Telescope (EHT) imaged the structure around the supermassive black hole Messier 87$^*$ (M87$^*$)~\cite{Akiyama:2019cqa}, finding a ring structure with an angular diameter of $42\pm3\mu as$, as well as a similar structure around the black hole at the center of our galaxy, Sagittarius A$^*$ (Sgr A$^*$)~\cite{2022ApJ...930L..12E}. This information, together with cosmological and weak field astrophysical data can be used to test different theories of gravity. 

A well-known approximation to study the deflection angle
near the photon sphere of a black hole was presented by Bozza in 2002~\cite{Bozza_2002}. Bozza's approximation separates and carefully describes the divergent part of the deflection angle at the photon sphere from the regular part. For a Schwarzschild black hole, Bozza's method can be compared with the exact result obtained by Darwin~\cite{doi:10.1098/rspa.1959.0015}
, giving a discrepancy in the deflection angle of about $0.06\%$. For other black holes, the results of the approximation are sometimes compared with the full numerical results, also giving good agreement near the photon sphere.  Several studies have used Bozza's approximation for deriving deflection angles predicted by alternatives to GR~\cite{Virbhadra:1998dy,Wei_2015,Zhao:2016kft,Badia:2017art,Izmailov:2019uhy}. Importantly, this method has also been applied to situations where the metric is only known numerically, as is the case in several modified gravity theories, and the corresponding results have been validated with full numerical computations of the deflection angle, which gives excellent agreement near the photon sphere~\cite{Chagoya:2020bqz}.

{In the literature, several works have investigated modifications to the Schwarzschild metric in order to impose bounds on the GUP parameters. However, the metric used in these works is not formally derived from the GUP principle but only motivated by it. For instance, in~\cite{scardigli2015gravitational}, upper limits 
for the GUP parameters were established from astronomical measurements, such as light deflection and perihelion 
precession, both for planets in the Solar System and for binary pulsars. Moreover, in~\cite{OKCU2021115324}, starting 
from the modified metric proposed in~\cite{scardigli2015gravitational}, the authors computed the Shapiro time delay, the gravitational 
redshift, and the geodetic precession. Based on observational data, they were able to place upper bounds on 
the GUP parameter. 
More recently, in~\cite{TURAKHONOV2024101716}, 
using the same effective metric proposed in~\cite{scardigli2015gravitational}, gravitational lensing in the strong field regime was 
investigated for GUP-modified Schwarzschild black holes. In that work, the authors assumed that the light ray 
follows a geodesic equation, an assumption supported by the results obtained in~\cite{PhysRevD.90.024023}, where the consistency 
of the equivalence principle within the GUP framework is demonstrated.}

In this work we apply Bozza's method to an effective static, spherically symmetric
black hole metric predicted in the context of {improved}-GUP, with the aim of obtaining precise calculations of the deflection angle near the photon sphere, where effects of the deformed metric, although still small, might be more relevant than in the asymptotically flat region of spacetime. Using current EHT data on observational properties of M87* and SgrA*, we derive constraints on the parameters of the modified metric. To our knowledge, these are the first constraints reported in the literature on the parameters of improved-GUP. 

This work is organized as follows. In Sec.~\ref{sec:The interior of Schwarzschild black hole}, we recast the Schwarzschild interior metric in terms of Ashtekar–Barbero variables, derive the classical equations of motion from the Hamiltonian constraint, and show how the interior and exterior geometries are recovered within this formalism. In Sec.~\ref{sec: Derivation of effective metric},  we derive the effective spacetime metric by implementing GUP- and EUP-inspired deformations of the classical algebra. We discuss the improved scheme, extend the interior solution to the full spacetime, and show that the resulting metric is regular everywhere and asymptotically flat. In Sec.~\ref{Sec:Deflectionangle}, we review Bozza's method for analyzing light deflection in the strong gravity regime. The approach isolates the logarithmic divergence near the photon sphere and provides a framework for computing the associated lensing observables.  In Sec.~\ref{Sec: Strong Lesing}, we analyze the strong lensing parameters derived from the black hole solutions and explore their observational implications. We compute how deviations from Schwarzschild values affect lensing observables such as the angular radius of the shadow and the separation between images, and confront these predictions with the cases of SgrA* and M87*. Finally, Sec.~\ref{Sec: Conclusions} is devoted to conclusions and final remarks.

\section{Schwarzschild black hole in Ashtekar-Barbero variables}
\label{sec:The interior of Schwarzschild black hole}
The Schwarzschild metric is an exact solution to the Einstein vacuum field equations that describes the gravitational field outside a spherically symmetric mass, assuming that the mass has no electric charge, no angular momentum, and that the cosmological constant is zero. In Schwarzschild coordinates, it is given by 
\begin{equation}
ds^{2}=-\left(1-\frac{R_s}{r}\right)dt^{2}+\left(1-\frac{R_s}{r}\right)^{-1}dr^{2}+r^{2}d\Omega^2,\label{eq:Usual_Schwarz_metric}
\end{equation}
with $r$ as the areal radial coordinate, $R_s = 2GM$ where $M$ is the asymptotic ADM mass and $G$ is the gravitational constant, and $d\Omega$ is the differential solid angle of a unit sphere. The Schwarzschild metric can also be interpreted as describing spacetime outside a static spherically symmetric black hole with event horizon at $R_s$. In this interpretation, the interior metric can be obtained by interchanging $t\leftrightarrow r$
\begin{equation}
ds^{2}=-\left(\frac{R_s}{t}-1\right)^{-1}dt^{2}+\left(\frac{R_s}{t}-1\right)dr^{2}+t^{2}d\Omega^2.\label{eq:sch-inter}
\end{equation}
Here, and throughout this work, $t$ is the Schwarzschild time coordinate with a range of $t\in(0,R_s)$ in the interior.

The metric Eq.~\eqref{eq:sch-inter} can also be expressed in terms of the Ashtekar–Barbero variables~\cite{Ashtekar_2006}, denoted by \( b \), \( c \), \( p_b \), and \( p_c \), where \( b \) and \( c \) are the configuration variables and \( p_b \) and \( p_c \) are their conjugate momenta. In this formalism, the metric takes the form
\begin{equation}
ds^{2}=-N(T)^{2}dT^{2}+\frac{p_{b}^{2}}{L_{0}^{2}|p_{c}|}dr^{2}+|p_{c}|d\Omega^2.\label{eq:Metric_Ashtkar_variables}
\end{equation}
The function $N$ is the lapse function that arises from the ADM (Arnowitt-Deser-Misner) decomposition of the interior spacetime. The timelike coordinate $T$ is a generic one that is associated with the choice of $N$, and is generally not the timelike coordinate of the Schwarzschild interior metric, which we call $t$. In addition, \( L_0 \) is a fiducial parameter introduced to make the symplectic structure, and thus the definition of the Poisson brackets, well defined. Clearly, no physical observables should depend on \( L_0 \) or on its rescaling. As we review below, metric Eq.~\eqref{eq:sch-inter} can be recovered from Eq.~\eqref{eq:Metric_Ashtkar_variables} by solving Hamilton's equations of motion for the Ashtekar-Barbero variables. The algebra of the canonical variables is 
\begin{align}
\{c,p_{c}\}= & 2G\gamma, & \{b,p_{b}\}= & G\gamma,\label{eq:classic-PBs-bc}
\end{align}
where $\gamma$ is the Barbero–Immirzi parameter~\cite{Thiemann_2007}, a free parameter in LQG typically assumed to be real and positive. 

The Hamiltonian constraint in the inner region of the Schwarzschild black hole is~\cite{Ashtekar_2006}
\begin{equation}
H=-\frac{8\pi N}{\gamma^2}\frac{\mathrm{sgn}(p_c)}{\sqrt{|p_{c}|}}\left[2bcp_c+\left(b^{2}+\gamma^{2}\right)p_{b}\right],\label{eq:H-class-N}
\end{equation}
where $\text{sgn}$ is the sign function. The set of canonical variables \( (b, p_b, c, p_c) \) forms a four-dimensional phase space \( \Gamma \), and the set of allowed points in \( \Gamma \) consists of those that satisfy the Hamiltonian constraint
\begin{equation}
\left(b^{2}+\gamma^{2}\right)\frac{p_{b}}{b}+2cp_{c}\approx 0.\label{eq:weak-van}
\end{equation}

The choice of an appropriate lapse function in Eq. \eqref{eq:H-class-N} allows for simplification of the Hamiltonian. There are several possible choices for $N$, for example, in \cite{Ashtekar_2006, Corichi_2016, https://doi.org/10.1155/2008/459290,Bosso_2021,PhysRevD.111.024048}, the following was considered
\begin{equation}
N\left(T\right)=\frac{\gamma\text{sgn}(p_c)\sqrt{|p_{c}\left(T\right)}|}{16\pi G b}.\label{eq:Lapse_funct_corichi_bosso}
\end{equation}
We adopt this choice since the equations of motion obtained for the pair $(c,\ p_c)$ decouple from those obtained for $(b,\ p_b)$ \footnote{Another choice of $N$ is found in \cite{PhysRevD.78.064040},
which is convenient for studying the dynamics of the Schwarzschild black hole in the context of LQG. Moreover, an alternative choice of the lapse function was recently employed in \cite{MELCHOR2024116584,Bosso_2024}, which allows for a simplified Hamiltonian suitable for quantization. Regarding physical outcomes, the choice of any lapse function mentioned here leads to the usual classical solution. Therefore, the equations of motion obtained from different choices of $N$ are equivalent.}. The Hamiltonian then becomes
\begin{equation}
H=-\frac{1}{2G\gamma}\left[\left(b^{2}+\gamma^{2}\right)\frac{p_b}{b} +2cp_{c}\right].\label{eq:H-class-1}
\end{equation}
From this Hamiltonian one can obtain the equations of motion for the dynamical variables $b(T)$, $p_b(T)$, $c(T)$, and $p_c(T)$. Using the classical Poisson algebra Eq.~\eqref{eq:classic-PBs-bc},
\begin{align}
\frac{db}{dT}= & \left\{ b,H\right\} =-\frac{1}{2}\left(b+\frac{\gamma^{2}}{b}\right),\label{eq:EoM-diff-b}\\
\frac{dp_{b}}{dT}= & \left\{ p_{b},H\right\} =\frac{p_{b}}{2}\left(1-\frac{\gamma^{2}}{b^{2}}\right),\label{eq:EoM-diff-pb}\\
\frac{dc}{dT}= & \left\{ c,H\right\} =-2c,\label{eq:EoM-diff-c}\\
\frac{dp_{c}}{dT}= & \left\{ p_{c},H\right\} =2p_{c},\label{eq:EoM-diff-pc}
\end{align}
the solution to this system of differential equations is 
\begin{align}
b\left(T\right)= & \pm\gamma\sqrt{2GMe^{-T}-1},\\
p_{b}\left(T\right)= & L_0 e^{\frac{T}{2}}\sqrt{2GM-e^{T}},\\
c\left(T\right)= & \mp\gamma GML_{0}e^{-2T},\\
p_{c}\left(T\right)= & e^{2T}.\label{eq:EoM-Sol-T-pc}
\end{align}
Here, the integration constants have been fixed in order to recover the spatial components of the interior metric Eq.~\eqref{eq:sch-inter}. The metrics Eq.~\eqref{eq:sch-inter} and Eq. \eqref{eq:Metric_Ashtkar_variables} describe the same spacetime and must therefore be related. In particular, there must exist a transformation relating the generic time parameter $T$ to the Schwarzschild time coordinate $t$, such that one recovers the interior metric components given in Eq.~\eqref{eq:sch-inter}. This relation is given by \( t = R_s e^{T} \)~\cite{Corichi_2016}. Thus, the solution to the equations of motion in terms of Schwarzschild time becomes
\begin{align}
b\left(t\right)= & \pm\gamma\sqrt{\frac{R_s}{t}-1},\label{eq:sol-cls-b}\\
p_{b}\left(t\right)= & L_{0}t\sqrt{\frac{R_s}{t}-1},\label{eq:sol-cls-pb}\\
c\left(t\right)= & \mp\frac{\gamma R_sL_{0}}{2t^{2}},\label{eq:sol-cls-c}\\
p_{c}\left(t\right)= & t^{2}.\label{eq:sol-cls-pc}
\end{align}
From these equations, one can see that $p_{c}\to0$ as $t\to0$, that is, at the classical singularity. 

Also, in Schwarzschild coordinates $(t,r,\theta,\phi)$, the general spherically symmetric metric corresponding to Eq.~\eqref{eq:Metric_Ashtkar_variables} is \cite{PhysRevD.111.024048}
\begin{equation}
ds^2 = -\frac{N(t)^2}{t^2} dt^2 + \frac{p_b^2(t)}{L_0^2 p_c(t)} dr^2 + p_c(t)d\Omega^2.\label{eq:Inter_metric}
\end{equation}
Since the extended (interior and exterior) metric of the classical Schwarzschild spacetime can be derived by switching the timelike and spacelike coordinates of the Schwarzschild interior, as a first attempt, we try to apply the same concept to the interior metric Eq. \eqref{eq:Inter_metric} to obtain the full spacetime metric. In other words, we interchange $t\leftrightarrow  r$ in the solutions Eqs.~\eqref{eq:sol-cls-b}--\eqref{eq:sol-cls-pc} and in the metric Eq.~\eqref{eq:Inter_metric}. Therefore, the extended (interior and exterior) spacetime metric is now \cite{PhysRevD.111.024048}
\begin{equation}
ds^2 = \frac{p_b^2(r)}{L_0^2 p_c(r)} dt^2 - \frac{1}{r^2} \frac{\gamma^2p_c(r)}{b^2(r)} dr^2 + p_c(r) d\Omega^2.\label{eq:Exterior_metric}
\end{equation}
From now on, we will consider both interior and exterior as described by the coordinates \((t, r, \theta, \phi)\), and we shall keep in mind that \(t\) and \(r\) are timelike and spacelike in the exterior, respectively, and spacelike and timelike in the interior, respectively.

\section{Derivation of the full spacetime effective metric}\label{sec: Derivation of effective metric}
In this section, we implement a modification to the classical algebra Eq.~\eqref{eq:classic-PBs-bc}, inspired by the minimal uncertainty approach. This framework is characterized by a modified commutation relation between position and momentum, which leads to a generalized version of the Heisenberg uncertainty principle. In particular, one such modification is the so-called Generalized Uncertainty Principle (GUP), which deforms the canonical commutation relation between the configuration operator $\hat{q}$ and its conjugate momentum $\hat{p}$ by introducing a quadratic correction in the momentum \cite{PhysRevD.52.1108}
\begin{eqnarray}
\left[\hat{q},\hat{p}\right]=i\hbar\left(1+\beta\  \hat{p}^2\right),\label{Eqq.GCR.1}
\end{eqnarray}
where $\beta$ is the deformation parameter. The generalized uncertainty relation associated with Eq.~\eqref{Eqq.GCR.1}  is
\begin{eqnarray}
    \Delta q\Delta p\geq\frac{\hbar}{2}\left(1+\beta\left(\Delta p\right)^2\right),\label{Eqq.GUP.1}
\end{eqnarray}
and it leads to a non-zero minimal uncertainty in position 
\begin{eqnarray}
    \Delta q_{\text{min}}=\hbar\sqrt{\beta},
\end{eqnarray}
which is associated with the quantization of spacetime~\cite{PhysRevD.52.1108}.

Similarly, there is another extension to the Heisenberg uncertainty relation called Extended  Uncertainty Principle (EUP) \cite{PhysRevD.79.125007, MUREIKA201988, doi:10.1142/S0217732319502043}, given by
\begin{eqnarray}
    \Delta q\Delta p\geq\frac{\hbar}{2}\left(1+\alpha\left(\Delta q\right)^2\right),\label{enq.GUP_q_p_min_p}
\end{eqnarray}
which leads to a non-zero minimal uncertainty in momentum \cite{PhysRevD.79.125007}
\begin{eqnarray}
     \Delta p_{\text{min}}=\hbar\sqrt{\alpha},\label{eq:Min_uncer_momentum}
\end{eqnarray}
and the modified commutation relation from which Eq. \eqref{enq.GUP_q_p_min_p} follows is expressed as
\begin{equation}
    \left[\hat{q},\hat{p}\right]=i\hbar\left(1+\alpha \hat{q}^2\right).\label{eq:Deformd_conmuta_relat_x_p}
\end{equation}
The corrections from EUP have gained relevance recently, as it was previously not believed to be necessary to introduce quantum corrections on a large scale in gravitational physics. This is no longer the case, and EUP provides a way to introduce quantum effects at macroscopic distances \cite{PhysRevD.79.125007, MUREIKA201988, doi:10.1142/S0217732319502043}.

Both Eq.~\eqref{Eqq.GCR.1} and Eq.~\eqref{eq:Deformd_conmuta_relat_x_p} serve different purposes and, consequently, lead to different predictions. In particular, in~\cite{Bosso_2020}, the approach of minimum uncertainty in the position is employed in quantum cosmology for a Kantowski-Sachs model describing the interior of a black hole. This new perspective yields a modified Wheeler-DeWitt equation, producing various outcomes compared to the conventional approach.

On the other hand, in works such as \cite{Bosso_2021,MELCHOR2024116584,Bosso_2024}, a modification to the classical algebra governing the interior dynamics of a Schwarzschild black hole was implemented. This modification is inspired by EUP and results in the resolution of the black hole's physical singularity, replacing it with a minimal area of quantum gravitational origin. 

In a recent study \cite{PhysRevD.111.024048}, it was shown that the extension of the modified interior metric obtained from the EUP prescription to the exterior spacetime leads to serious problems in the asymptotic region of the exterior. To address this, the authors introduce an “improved scheme” that mimics a similar prescription used in LQG, where quantum parameters are promoted to momentum-dependent functions~\cite{PhysRevD.78.064040,PhysRevD.74.084003,chiou2013loopquantizationsphericallysymmetric}.

The effective algebra proposed in~\cite{Bosso_2021}, as a modification of Eq.~\eqref{eq:classic-PBs-bc} within the minimal uncertainty framework, is given by
\begin{align}
\{b, p_b\} &= G\gamma(1 + \beta_b b^2), \label{eq:Mod_b_pb_algeb} \\
\{c, p_c\} &= 2G\gamma(1 + \beta_c c^2), \label{eq:Mod_c_pc_algeb}
\end{align}
where \(\beta_b\) and \(\beta_c\) are small parameters that, in this case, represent the effective deformation parameters that quantify the modifications introduced in Eq.~\eqref{eq:Mod_b_pb_algeb} and Eq. \eqref{eq:Mod_c_pc_algeb}. These effects are expected to be appreciable in the semiclassical regime as Planck-scale suppressed corrections, although the interpretation of such modifications may be more subtle than in the quantum regime~\cite{Bosso_2023}.

From the effective dynamics arising from the modified classical algebra, given in Eq.~\eqref{eq:Mod_b_pb_algeb} and Eq.~\eqref{eq:Mod_c_pc_algeb}, one can derive the components of the metric that incorporate effective corrections. In the next section, we present the derivation of the effective metric considering that the deformation parameters are momentum-dependent~\cite{PhysRevD.111.024048}. An alternative metric, which is not relevant for our work, can be derived when the deformation parameters are constants \cite{Bosso_2021}.



\subsection{Improved metric and its extension to  full spacetime}\label{Subsc:A_Improved_GUP}
As mentioned previously, the modification of the classical algebra, shown in Eq.~\eqref{eq:Mod_b_pb_algeb} and Eq.~\eqref{eq:Mod_c_pc_algeb}, was implemented in~\cite{Bosso_2021,MELCHOR2024116584} by keeping the deformation parameters $\beta_b$ and $\beta_c$ constant. This was done with the aim of resolving, at least effectively, the physical singularity located in the black hole interior. Applying this deformation to the interior dynamics of the black hole leads to modifications in the components of the metric. However, it was recently shown in \cite{PhysRevD.111.024048} that this metric presents certain problems, mainly in the asymptotic limit \( r \to \infty \). To address these problems, the same work proposed an improved technique that solves them, which we briefly discuss below.

Following the procedure developed in \cite{PhysRevD.111.024048}, we promote the effective parameters $\beta_b$ and $\beta_c$ in Eq.~\eqref{eq:Mod_b_pb_algeb} and Eq.~\eqref{eq:Mod_c_pc_algeb} to momentum-dependent functions. This can be achieved by redefining these parameters as
\begin{align}
\beta_b &\rightarrow \bar{\beta}_b = \frac{\beta_b L_0^4}{p_b^2},\label{eq:Improv_beta_b} \\
\beta_c &\rightarrow \bar{\beta}_c = \frac{\beta_c L_0^4}{p_c^2}, \label{eq:Improv_beta_c}
\end{align}
where the powers of $L_0$ are included to render $\bar{\beta}_{b}$ and $\bar{\beta}_{c}$ dimensionless. This prescription is named the $\bar{\beta}$-improved scheme~\cite{PhysRevD.111.024048}. As a result of the above, the effective GUP-induced algebra now becomes
\begin{align}
\{b, p_b\} &= G\gamma(1 + \bar{\beta}_b b^2) = G\gamma\left(1 + \frac{\beta_b L_0^4}{p_b^2} b^2\right), \label{eq:Improv_mod_algebra_b_pb} \\
\{c, p_c\} &= 2G\gamma(1 + \bar{\beta}_c c^2) = 2G\gamma\left(1 + \frac{\beta_c L_0^4}{p_c^2} c^2\right). \label{eq:Improv_mod_algebra_c_pc}
\end{align}
Implementing these new algebras in the interior dynamics of the black hole leads to modified equations of motion~\cite{PhysRevD.111.024048}, given by
\begin{align}
\frac{db}{dT}= & \left\{ b,H\right\} =-\frac{1}{2b}{\left(b^2+\gamma ^2\right) \left(\frac{b^2 \beta_b L_0^4}{p_b^2}+1\right)},\label{eq:Improv_EoM-diff-b}\\
\frac{dp_{b}}{dT}= & \left\{ p_b,H\right\} =\frac{1}{2 b^2 p_b} {\left(b^2-\gamma ^2\right) \left(\beta_b L_0^4b^2+p_b^2\right)},\label{eq:Improv_EoM-diff-pb}\\
\frac{dc}{dT}= & \left\{ c,H\right\} = -2 c \left(\frac{\beta_c L_0^4c^2}{p_c^2}+1\right),\label{eq:Improv_EoM-diff-c}\\
\frac{dp_{c}}{dT}= & \left\{ p_c,H\right\} =2 p_c \left(\frac{\beta_c c^2 L_0^4}{p_c^2}+1\right).\label{eq:Improv_EoM-diff-pc}
\end{align}
Following a procedure analogous to that employed in the previous section, the modified equations of motion can be solved and rewritten in terms of \( t \), using the relation \( t = R_s e^{T} \). The resulting solution takes the form
\begin{align}
b =& \gamma \sqrt{\frac{R_s}{\sqrt{t^2-\bar{Q}_b}}-1}, \label{eq:Improv_b_solut} \\
p_b =& L_0 \sqrt{t^2 - \bar{Q}_b}\sqrt{\frac{R_s}{\sqrt{t^2-\bar{Q}_b}}-1}, \label{eq:Improv_pb_solut} \\
c =& -\frac{\gamma L_0 R_s}{2} \left(t^8 - \frac{1}{4}\bar{Q}_cR_s^2\right)^{-\frac{1}{4}}, \label{eq:Improv_c_solut} \\
p_c =& \left( t^8 - \frac{1}{4}\bar{Q}_cR_s^2 \right)^{\frac{1}{4}}. \label{eq:Improv_pc_solut}
\end{align}
In the above equations, we have defined two dimensionful effective parameters, given by
\begin{align}
\bar{Q}_b & = \text{sgn}(\beta_b)|\beta_b| \gamma^2 L_0^2, \quad  \quad \bar{Q}_c = \text{sgn}(\beta_c)|\beta_c| \gamma^2 L_0^6, \label{eq:Improv_parameters_Qb_Qc}
\end{align}
where $\bar{Q}_b$ has dimensions of $[L_0]^2$ and $\bar{Q}_c$ has dimensions of $[L_0]^6$.

As in the standard classical Schwarzschild case, we postulate that the full spacetime metric of this GUP-inspired black hole can be obtained by analytically extending the interior solution through the coordinate transformation $t \leftrightarrow  r$. To construct the full spacetime metric in this manner, we apply the aforementioned coordinate relabeling to Eqs. \eqref{eq:Improv_b_solut}--\eqref{eq:Improv_pc_solut}, yielding the extended expressions for the canonical variables as \cite{PhysRevD.111.024048}
\begin{align}
b &= \gamma \sqrt{\frac{R_s}{\sqrt{r^2 - \bar{Q}_b}} - 1}, \label{eq:Improv_solut_b} \\
p_b &= L_0 \sqrt{r^2 - \bar{Q}_b} \sqrt{\frac{R_s}{\sqrt{r^2 - \bar{Q}_b}} - 1}, \label{eq:Improv_solut_pb} \\
c &= -\frac{\gamma L_0 R_s}{2 } \left(r^8 - \frac{1}{4}\bar{Q}_cR_s^2\right)^{-\frac{1}{4}}, \label{eq:Improv_solut_c} \\
p_c &= \left(r^8 - \frac{1}{4}\bar{Q}_cR_s^2\right)^{\frac{1}{4}}. \label{eq:Improv_solut_pc}
\end{align}
In order to ensure that the metric remains real for all values of the coordinates, both 
\( \bar{Q}_b \) and \( \bar{Q}_c \) in Eqs.~\eqref{eq:Improv_solut_b}--\eqref{eq:Improv_solut_pc} 
must be negative~\cite{PhysRevD.111.024048}. Consequently, we infer that \( \text{sgn}(\beta_b) = -1 \) and \( \text{sgn}(\beta_c) = -1 \). Therefore, we rewrite 
\(\bar{Q}_b=-|\bar{Q}_b|\) and \(\bar{Q}_c=-|\bar{Q}_c|\), with 
\(|\bar{Q}_b|=\ell_{{b}}^2\gamma^2\) and \(|\bar{Q}_c|=\ell_{{c}}^6\gamma^2\). 
Here, we have defined \( \ell_{{b}}^2=\beta_b L_0^2 \) and \( \ell_{{c}}^6=\beta_c L_0^6 \) 
as two fundamental physical minimum length scales of the theory.
Substituting the solutions Eqs.~\eqref{eq:Improv_solut_b}--\eqref{eq:Improv_solut_pc} into the full spacetime metric Eq.~\eqref{eq:Exterior_metric}, we obtain the components of the improved full metric
\begin{multline}
    g_{00} =  -\left(1+\frac{|\bar{Q}_b|}{r^2}\right)\left(1+\frac{|\bar{Q}_c| R_s^2}{4r^8}\right)^{-1/4}\\
      \times \left(1-\frac{R_s}{\sqrt{r^2+|\bar{Q}_b|}}\right), \label{eq:Improv_goo_compont}
\end{multline}
\begin{align}
g_{11} &=  \left(1+\frac{|\bar{Q}_c| R_s^2}{4r^8}\right)^{1/4}\left(1-\frac{R_s}{\sqrt{r^2+|\bar{Q}_b|}}\right)^{-1}, \label{eq:Improv_g11_compont} \\
g_{22} &= \frac{g_{33}}{\sin^2\theta} =  r^2\left(1+\frac{|\bar{Q}_c| R_s^2}{4r^8}\right)^{1/4}. \label{eq:Improv_g22_g33_compont}
\end{align}
The coordinates have the domain $r \in [0, +\infty)$, $t \in (-\infty, +\infty)$, $\theta \in [0, \pi]$, and $\phi \in [0, 2\pi)$.

We now analyze the behavior of the effective components, Eqs.~\eqref{eq:Improv_goo_compont}--\eqref{eq:Improv_g22_g33_compont}, derived within the improved GUP scheme, in three regions of interest: \( r \to 0 \), \( r \to R_s \), and \( r \to \infty \). In the first region, $r\to0$, we find that these effective components reduce to
\begin{align}
 \lim_{r \to 0} g_{00}&=\frac{\sqrt{2} \left(\sqrt{|\bar{Q}_b|} R_s-|\bar{Q}_b|\right)}{\left(|\bar{Q}_c| R_s^2\right)^{1/4}},\label{eq:Improve_g_00_r_to_0}\\
 \lim_{r \to 0} g_{11} &\to\infty, \label{eq:Improve_g_11_r_to_0}\\
 \lim_{r \to 0} g_{22} &= \frac{g_{33}}{\sin^2\theta} =\frac{\left(|\bar{Q}_c| R_s^2\right)^{1/4}}{\sqrt{2}}.\label{eq:Improve_g_22_g_33_r_to_0}
\end{align}
It is clear that the \( g_{00} \) component of the standard metric in Eq.~\eqref{eq:Usual_Schwarz_metric} differs significantly from that in Eq.~\eqref{eq:Improve_g_00_r_to_0} in the limit $ r \to 0 $, where the former diverges. Interestingly, the only feature that persists is the divergence of \( g_{11} \) in the region \( r \to 0 \). However, this corresponds to a coordinate singularity rather than a physical one, as can be confirmed by evaluating the Kretschmann scalar, which
in the limit \( r \to 0 \) approaches a finite, positive, and nonzero value given by \cite{PhysRevD.111.024048}
\begin{equation}
\lim_{r \to 0}K =\frac{8}{R_{s}\sqrt{|\bar{Q}_c|}}
\end{equation}
which provides evidence that no physical singularity is present in this region. 
Likewise, if \( \beta_c \to 0 \), the classical singularity reappears within the improved-GUP framework. 
It should be noted that the resolution of the singularity is governed solely by the presence of the effective parameter \( \bar{Q}_c \).

In the region \( r = R_s \), the effective components given in Eqs.~\eqref{eq:Improv_goo_compont}--\eqref{eq:Improv_g22_g33_compont} take the following values
\begin{align}
    \lim_{r \to R_s} g_{00}&=-\frac{\left(1+\frac{|\bar{Q}_b|}{R_s^2}\right) \left(1-\frac{R_s}{\sqrt{|\bar{Q}_b|+R_s^2}}\right)}{\left(\frac{|\bar{Q}_c|}{4 R_s^6}+1\right)^{1/4}},\\
    \lim_{r \to R_s} g_{11}&=\frac{\left(\frac{|\bar{Q}_c|}{4 R_s^6}+1\right)^{1/4}}{1-\frac{R_s}{\sqrt{|\bar{Q}_b|+R_s^2}}},\\
    \lim_{r \to R_s} g_{22}&=\frac{g_{33}}{\sin^2\theta} =R_s^2 \left(\frac{|\bar{Q}_c|}{4 R_s^6}+1\right)^{1/4}.
\end{align}
On the other hand, in the asymptotic limit \( r \to \infty \), we have
\begin{align}
    \lim_{r \to \infty} g_{00}&=-1,\label{eq:asymp00}\\
    \lim_{r \to \infty} g_{11}&=1, \label{eq:asymp11}\\
    \lim_{r \to \infty} g_{22}&\to r^2 \label{eq:asymp2233}.
\end{align}
This implies that the effective metric components derived within the improved-GUP framework reduce to those of flat spacetime in the asymptotic limit. This not only resolves the issues encountered with the components obtained in the unimproved GUP approach, as derived in ~\cite{Bosso_2021,PhysRevD.111.024048}, but also eliminates the physical singularity that was present at the classical level in \( r = 0 \).

From the effective \( g_{00} \) component of the metric in Eq.~\eqref{eq:Improv_goo_compont}, 
the event horizon can be determined by defining a radius of the horizon \( R_H \) such that \( g_{00}(R_H)=0 \). 
This is possible because, in a spherically symmetric and static spacetime, the event horizon 
coincides with the Killing horizon, as also pointed out in~\cite{PhysRevD.111.024048}. 
Accordingly, the effective horizon radius is
\begin{equation}
    R_H=R_s\sqrt{1-\frac{|\bar{Q}_b|}{R_s^2}}.\label{eq_Eff_Rad_Horz}
\end{equation}
As can be seen from this expression, the horizon radius \( R_H \) is smaller than the Schwarzschild 
radius \( R_s \) by an amount which, although small, may play a significant role in black hole phenomenology.

\begin{figure}
   \centering
\includegraphics[width=0.42\textwidth]{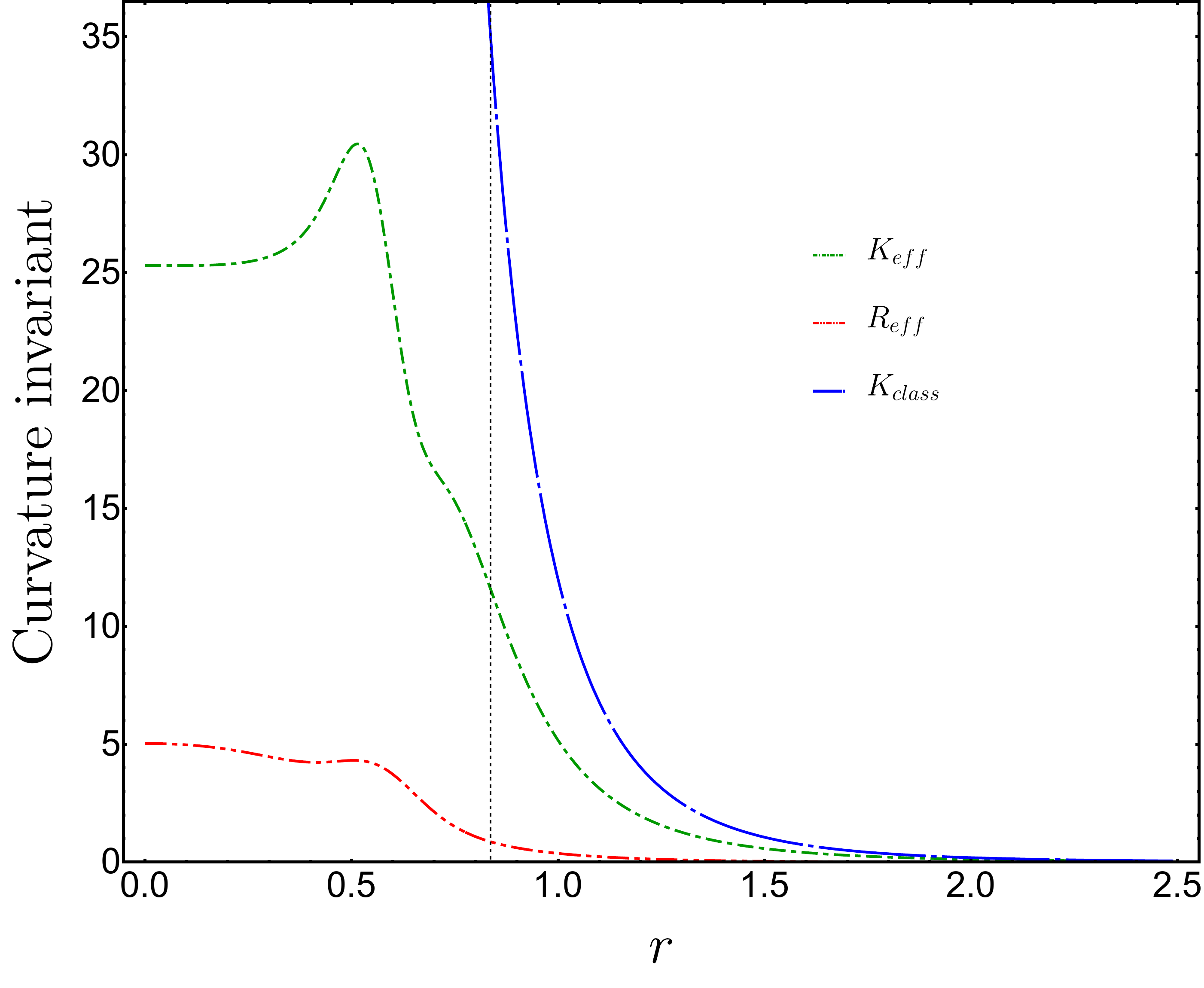}
  \caption{\it{Plots of Ricci and Kretschmann scalars corresponding to the improved metric, whose components are given by Eqs.~\eqref{eq:Improv_goo_compont}--\eqref{eq:Improv_g22_g33_compont}, as well as the Kretschmann scalar corresponding to the usual Schwarzschild metric, denoted by $K_{class}.$  We note that both scalars corresponding to the improved metric reach a finite value at $r=0$, while the classical Kretschmann scalar is divergent at $r=0$; the  classical  Ricci scalar is zero. The vertical dashed line indicates the position of the modified  horizon which is determined by $R_{H}=\sqrt{R_{s}^2-|\tilde{Q}_b|}$.  It is worth mentioning that the Ricci scalar has units of $[\text{Length}]^{-2}$, while the Kretschmann scalar has units of $[\text{Length}]^{-4}$}. For all curves we used $|\tilde{Q}_b|=0.3$, $|\tilde{Q}_c|=3$ and $R_{s}=1$.}
  \label{fig:invariant}
\end{figure}

In Fig.~\ref{fig:invariant}, we show the Ricci and Kretschmann scalars for an arbitrary choice of parameters $|\bar{Q}_b|$ and $|\bar{Q}_c|$ compared to the classical curvature scalars. Although in this plot the Ricci scalar of the improved metric is always positive for the chosen values of the parameters, it is possible to find values such that the Ricci scalar may develop a transition from positive to negative curvature, followed by the expected asymptotically flat behavior. This suggests that the improved GUP corrections to the metric alter the spacetime geometry in a non-trivial way, which may behave as an effective source of negative curvature. 

From this point onward, we can work with a metric that is regular throughout the entire spacetime, incorporates effective quantum gravity corrections, and reduces to a flat metric in the asymptotic limit. What remains to be understood is how significant these corrections are and how they affect light deflection. To address these questions, in the following section we compute the deflection angle using the effective metric components given in Eqs.~\eqref{eq:Improv_goo_compont}--\eqref{eq:Improv_g22_g33_compont}.


\section{Deflection angle in the strong gravity regime} \label{Sec:Deflectionangle}
Let us briefly review the analytical method developed by Bozza~\cite{Bozza_2002} to calculate the deflection angle in the strong gravity regime. A more detailed explanation can be found in~\cite{Chagoya:2020bqz}. We consider geometries described by the line element
\begin{equation}
ds^2 = -A(r)dt^2 + B(r) dr^2 + C(r) d\Omega^2\,.
\label{eq:lineel}
\end{equation}
Bozza's approximation is valid when the metric components $A,B,C$ 
%
fulfill the asymptotic conditions $\lim_{r\to\infty}A =\lim_{r\to\infty}B = 1 $ and $\lim_{r\to\infty}C/r^2 = k$ for any constant $k$.

\begin{figure}
\includegraphics[width=0.42\textwidth]{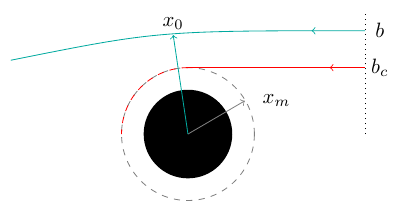}
\caption{{\it Schematic diagram for the deflection of light by a black hole. Photons approaching with the critical impact parameter $b_c$ enter a circular unstable orbit  with radius $x_m$, the radius of the photon sphere. } }
\label{ld}
\end{figure}
Strong lensing (see Fig.~\ref{ld} for a schematic representantion) refers to the deflection of light near the photon sphere, {i.e.}, the surface of radius $x_m$ given by the outermost solution of the equation
\begin{equation}
\frac{C'(x)}{C(x)} = \frac{A'(x)}{A(x)}\,,\label{eq:xm}
\end{equation}
where primes denote derivatives with respect to $x$. 
{Equation~\eqref{eq:xm} arises from requiring that the deflection angle in the spacetime described by Eq.~\eqref{eq:lineel} becomes unboundedly large at the photon sphere~\cite{Virbhadra:2002ju}.} In Eq.~\eqref{eq:xm}, 
following the classical computation of~\cite{Weinberg:100595}, it has been shown~\cite{Virbhadra:1998dy} that for the metric~\eqref{eq:lineel}  the deflection angle of photons near the photon sphere can be approximated as
\begin{equation}
\alpha(b) = -c_1 \log\left(\frac{b}{b_c} -1 \right) + c_2 + \mathcal O[(b-b_c)\log(b -b_c)]\,, \label{eq:defabeta}
\end{equation}
where $b$ denotes the impact parameter\footnote{Not to be confused with the dynamical variable $b$ defined in Eq.~\eqref{eq:classic-PBs-bc}. From now on, $b$ denotes the impact parameter.} of a light ray with distance of closest approach $x_0$,\footnote{{This result can be derived from normalization of the photon's 4-velocity and the conservation of angular momentum~\cite{Bozza_2002}.}}
\begin{equation}
b(x_0) = \sqrt{\frac{C(x_0)}{A(x_0)}}\,,\label{eq:impact}
\end{equation}
$b_c$ corresponds to $x_0=x_m$, and the coefficients $c_1, c_2$ are given by 
\begin{align}
c_1 &= \sqrt{\frac{2 A_m B_m}{ C''_m A_m - C_m A''_m}}\,,\label{eq:c1}
\\
c_2 &= c_1 \log\left[x_m^2 \left( \frac{C_m''}{C_m} - \frac{A_m''}{A_m}\right) \right] + I_R(x_m) - \pi\,,\label{eq:c2}
\end{align}
where a subscript $m$ denotes evaluation at the photon sphere, $x_m$, and
\begin{align}
I_R(x_m) = & 2 x_m \int_0^1 dz \left[ \sqrt{\frac{B}{C}}\left(\frac{C\,A_m}{A\, C_m} - 1 \right)^{-\frac{1}{2}} \frac{1}{(1-z)^2}\right. \nonumber \\ 
&\left. - \frac{c_1}{x_m\, z}\right]\,
\label{eq:ir}
\end{align}
is a regular integral. The divergent part of the deflection angle near the photon sphere has been singled out, it is given by the logarithmic term in Eq.~\eqref{eq:defabeta}. Notice that the integration variable in $I_R$ has been defined implicitly as  
\begin{equation}
x(z) = \frac{x_m}{1-z}\, ,\label{eq:zdef}
\end{equation}
which is introduced to avoid integrating up to infinity.

The coefficients $c_1$ and $c_2$ allow for a simple computation of two lensing observables: the magnification ratio 
\begin{align}
    r_\mu \approx e^{2\pi/c_1}, 
\end{align}
{which is the ratio between the magnification of the outermost image located at an angular position $\theta_1$ and the
sum of the magnifications of all the other images, whose position quickly approaches 
$\theta_\infty$}, and the separation between the first relativistic image and the others, given by
\begin{align}
s = \theta_1 - \theta_\infty & \approx 
\theta_{\infty} \exp\left(  \frac{c_2 - 2\pi}{c_1}\right) \nonumber \\
& = 
\frac{b_c}{D_{OL}} \exp\left(  \frac{c_2 - 2\pi}{c_1}\right)\,,\label{eq:sep}
\end{align}
where $D_{OL}$ is the distance between the observer and the lens. A detailed derivation of these observables can be found in~\cite{Chagoya:2020bqz}. 
The observables $r_\mu$  and $s$ are completely determined by $c_1$, $c_2$, $b_c$
and $D_{OL}$. This last quantity is fixed by observations~\footnote{A variety of observations are used for estimating the distance to astronomical objects. For local objects, geometric estimations based on orbital measurements are available, while for more distant objects astronomers resort to standard candles.},  while the other three
depend on the parameters that appear in the black hole solution under consideration, i.e., on the mass -- also observed -- and on the parameters that appear as a result of
considering alternative models of gravity.
In the next section, we present results for the quantities $c_1$ and $c_2$, as well as for the observables computed with these quantities, i.e., the deflection angle, the magnification ratio and the separation between images. 

\section{Strong lensing parameters and Observational constraints}\label{Sec: Strong Lesing}
In this section, we apply Bozza's method to our improved metric whose components are given by Eqs.\eqref{eq:Improv_goo_compont}-\eqref{eq:Improv_g22_g33_compont}, and, in order to use the notation of Sec.~\ref{Sec:Deflectionangle} we identify $g_{00}=A(r),\ g_{11}=B(r),\ g_{22}=C(r)$ and $g_{33}=g_{22}\sin^2\theta$. Clearly, these metric components satisfy the conditions required in Bozza's method, this can be explicitly noticed from Eqs.~\eqref{eq:asymp00}-\eqref{eq:asymp2233}. For dimensional analysis purposes, we assign the values $|\tilde{Q}_b| = |\bar{Q}_{b}|/ R_{s}^{2}$, $|\tilde{Q}_c| = |\bar{Q}_{c}|/ R_{s}^{6}$, and $R_{s}=2GM.$

\begin{figure}    \includegraphics[width=0.45\textwidth]{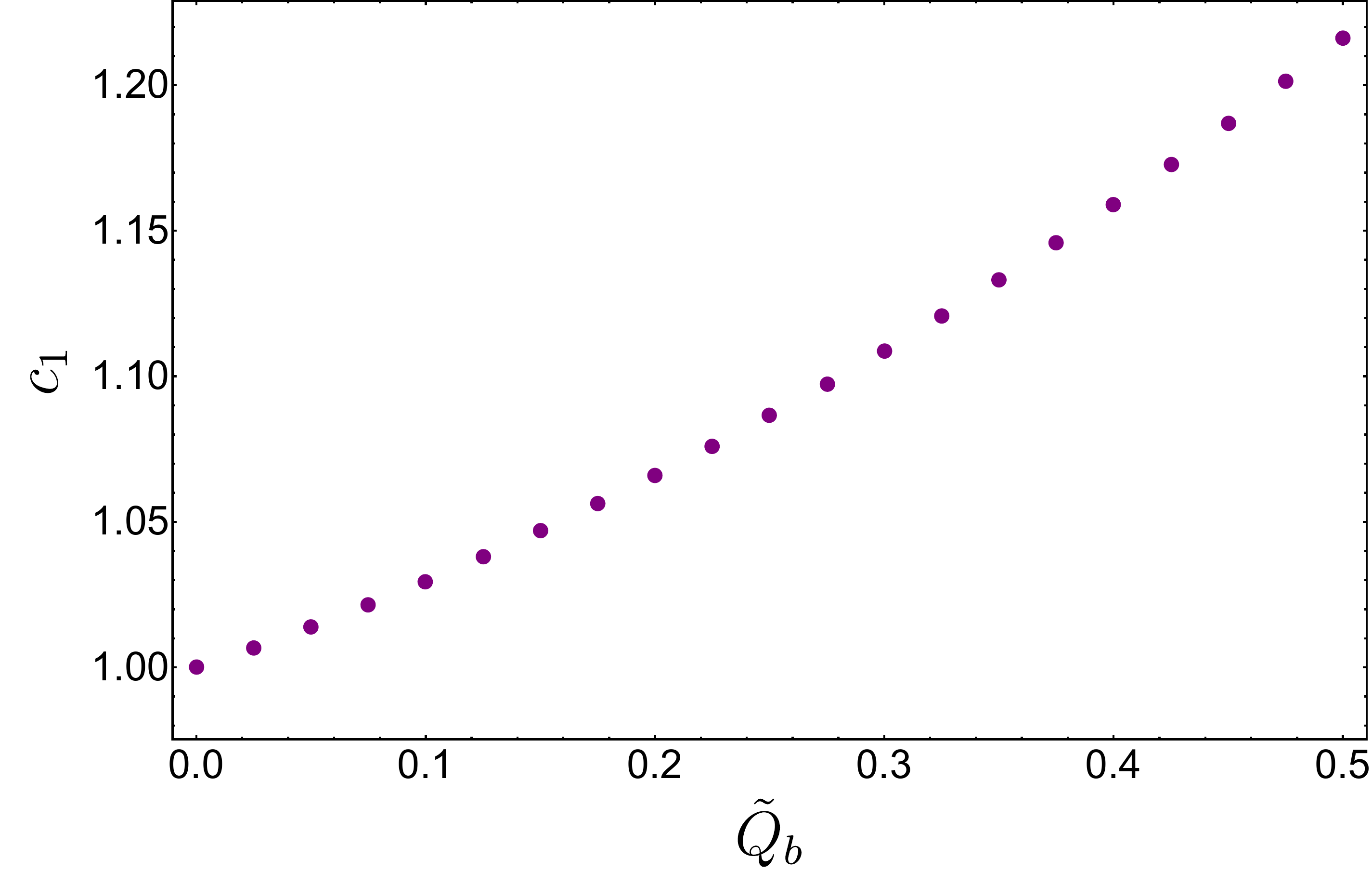}
    \includegraphics[width=0.45\textwidth]{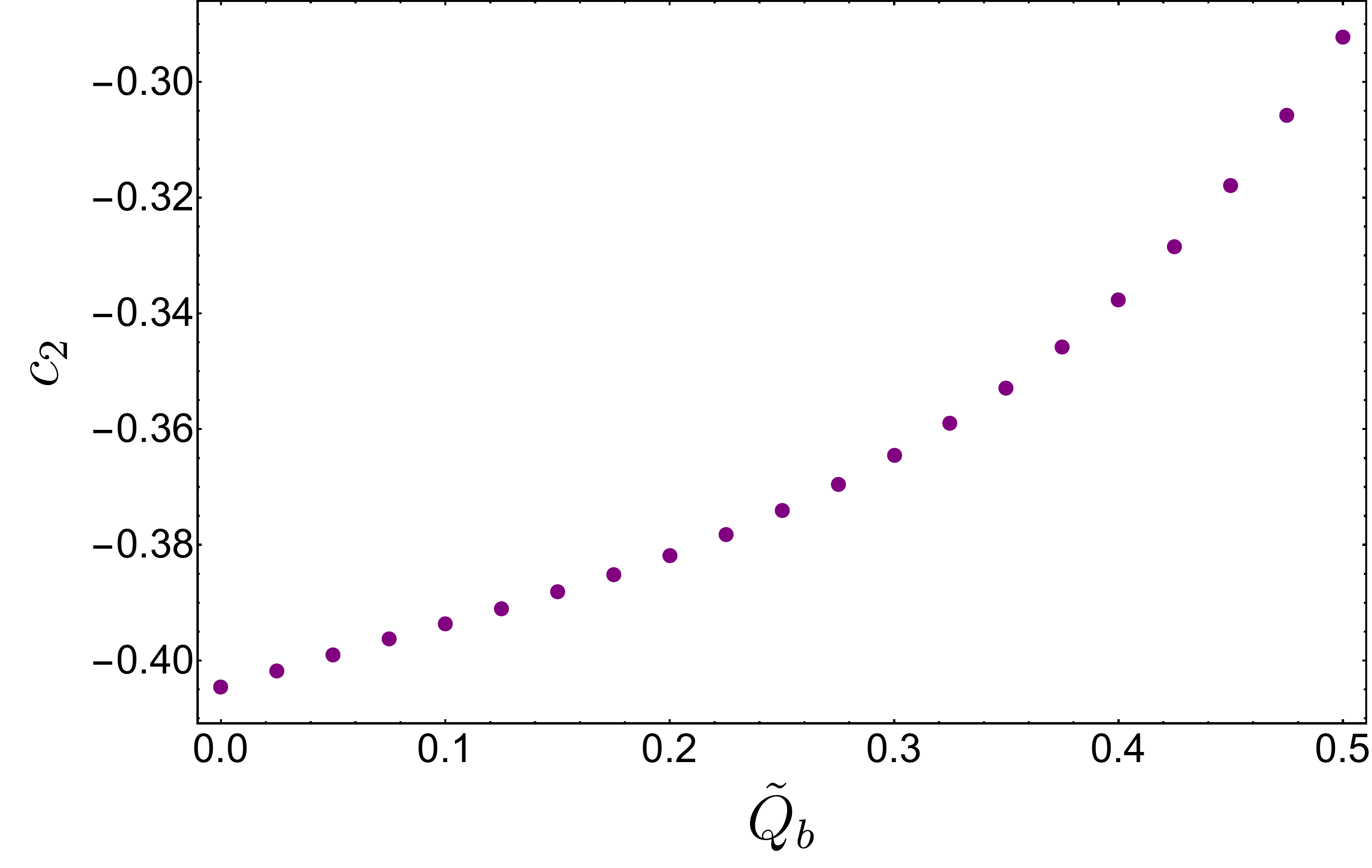}
    \caption{\it{Parameters $c_1$ and $c_2$ that determine the strong deflection limit in Bozza's approximation, for $0\leq|\tilde{Q}_b|\leq0.5$, and $|\tilde Q_c|=10^{-1}$.}}
    \label{fig:c1c2}
 \end{figure}

{As can be seen from Eq.~\eqref{eq:c1} and Eq.~\eqref{eq:c2}}, the coefficients $c_1$ and $c_2$ are computed purely from the metric components, thus, using the variable $x=r/R_s$ and the constants $|\tilde Q_b|, |\tilde Q_c|$, it is found that $c_1, c_2$ do not depend explicitly on $R_s$ but only on $|\tilde Q_b|$ and $|\tilde Q_c|$. This dependence is illustrated in Fig~\ref{fig:c1c2}. Both $c_1$ and $c_2$ have higher values than in the Schwarzschild case, which is approximated when $|\tilde Q_b|\to 0$. The magnitude of the changes in $c_1$ and $c_2$ is comparable, for example, to the variation induced by the electric charge in Reissner–Nordström (RN) black holes approaching the extremal limit, although the overall behavior of these parameters differs quantitatively from that of RN and other black holes.

\begin{figure}
\includegraphics[width=0.45\textwidth]{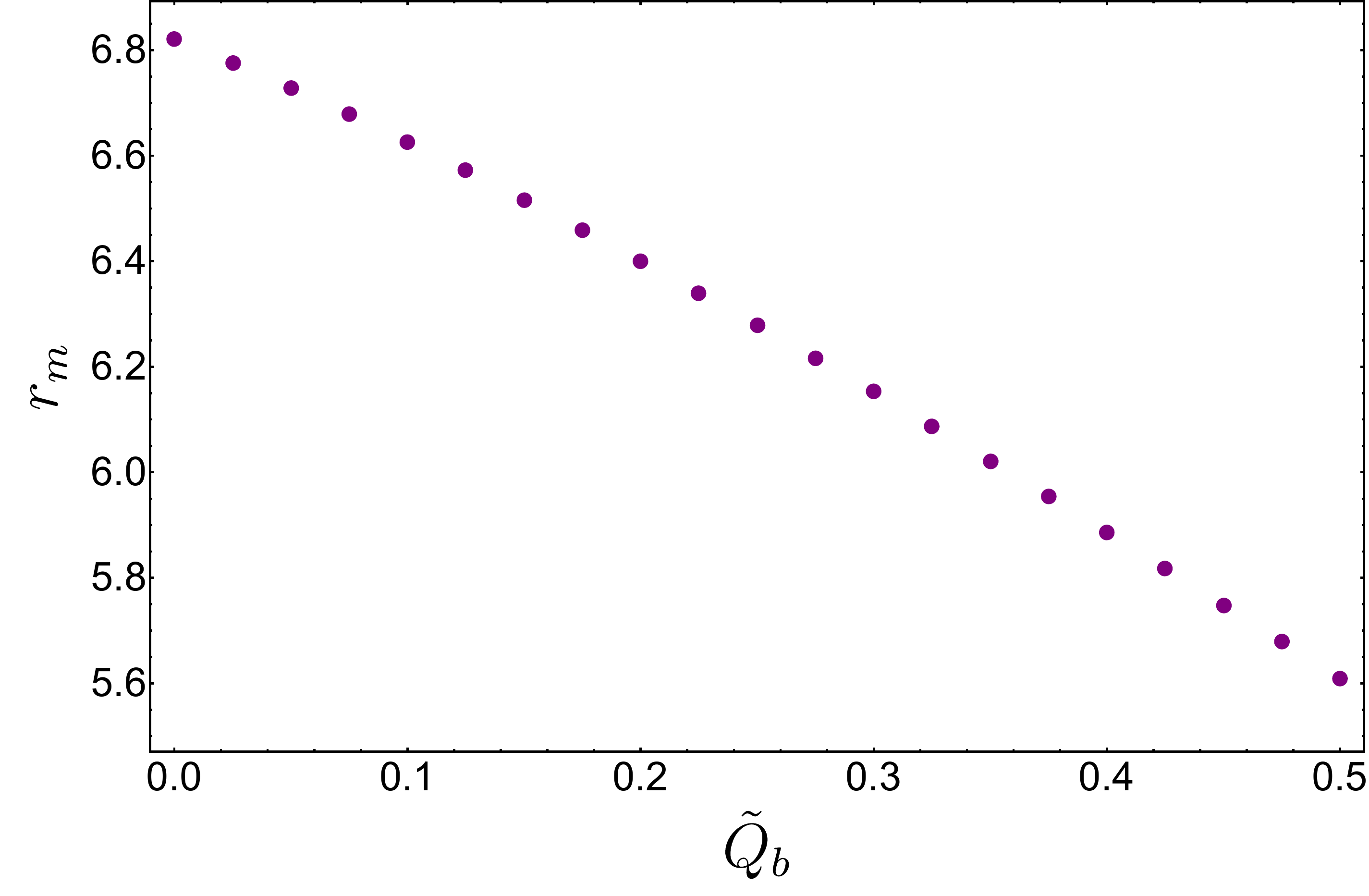}
    \includegraphics[width=0.45\textwidth]{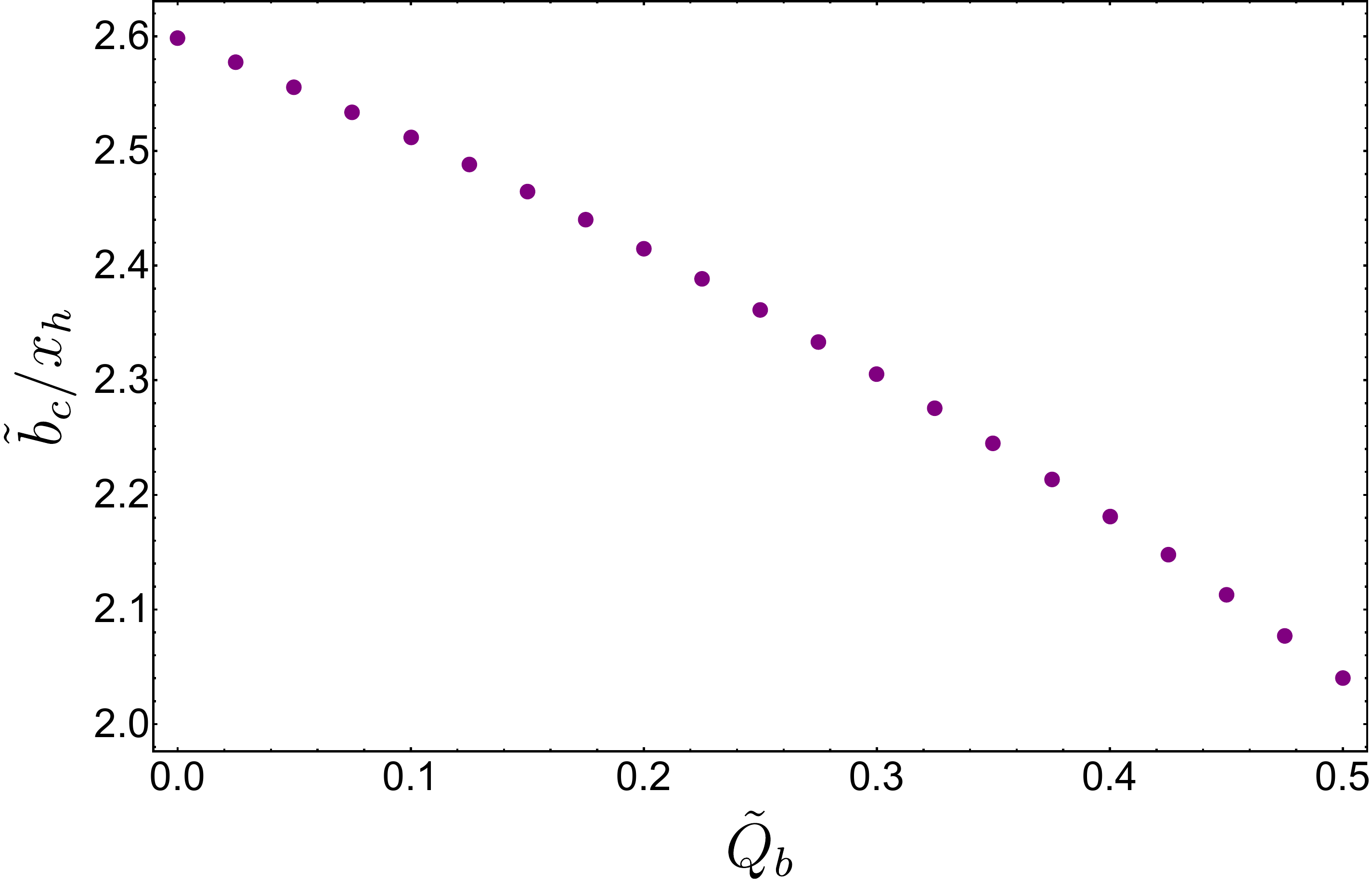}
  \caption{\it{Critical impact parameter, $\tilde b_c = b_c/R_s$, divided by the
    horizon of each solution, for $0\leq|\tilde Q_b|\leq0.5$.}}
  \label{fig:def2}
\end{figure}

 In the upper panel of Fig.~\ref{fig:def2} we show the magnification ratio as $r_m=2.5\log_{10} r_\mu$. We express it in this form in order to agree with the conventions used in the literature. As expected from the results for $c_1$, the magnification ratio behaves qualitatively similar to the case of an RN black hole (see Table I of~\cite{Bozza_2002}), with the difference that instead of an electric charge we are varying the parameter $|\tilde Q_b|$.

In the bottom panel of Fig.~\ref{fig:def2} we show
the ratio between the critical
impact parameter and the black hole horizon for the different solutions corresponding to different values of $|\tilde Q_b|$. We see that these black holes are weaker than a Schwarzschild black hole, in the sense that their capture radius for photons becomes smaller in proportion to their horizon radius.

\begin{figure}
   \centering
\includegraphics[width=0.45\textwidth]{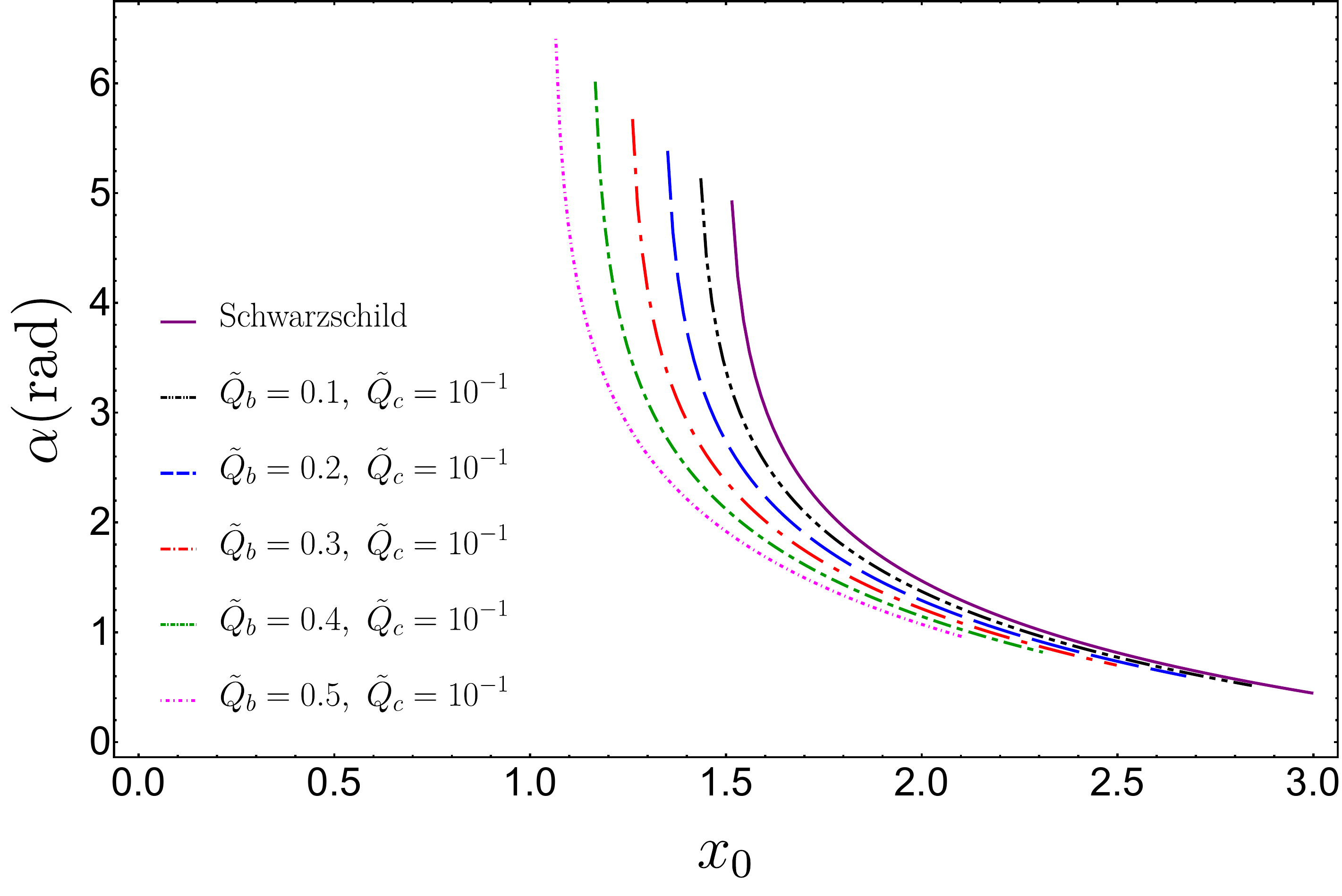}
  \caption{\it{Deflection angle as a function of $x_{0} = r_0/{2M}$. The dashed line represents the effective black hole, while the solid purple line corresponds to the exact Schwarzschild solution. As $|\tilde{Q}_{b}|$ increases, the deflection angle $\alpha$ deviates from its Schwarzschild value, gradually moves further to the left.}}
  \label{fig:def1}
\end{figure}

The results shown in Figs.~\ref{fig:c1c2} and~\ref{fig:def2} give us some insight into the lensing properties of the black holes under study. Additional information is provided by the observables $\theta_\infty$ and $s$, that is, the angular radius of the shadow and the separation between images.  These quantities depend on the distance between the observer and the lens and on the lens mass, which are determined observationally. In Tables~\ref{table:sgr}
 and~\ref{table:m87} we present the results for the supermassive black holes M87$^*$ and Sgr A$^*$, respectively. The masses and distances are taken from~\cite{Akiyama:2019eap} and~\cite{Gillessen_2017}. In both cases, we see that $|\tilde Q_b| > 0.3$ is in tension with
 EHT observations of $\theta_\infty$, which have an uncertainty of around $3\mu as$.

 Finally, in Fig.~\ref{fig:def1} we show the deflection angle for different values of $|\tilde Q_b|$ and a fixed $|\tilde Q_c| = 10^{-1}$, as a function of the distance of closest approach, $x_0$. For comparison, we also show the results for the Schwarzschild spacetime. We observe that all curves rapidly converge to the Schwarzschild results as the radius increases. This confirms that the region near the photon sphere is a better laboratory to test for lensing effects due to the improved-GUP spacetime.
 
\renewcommand{\arraystretch}{1.2}
\begin{table}
\caption{\label{table:sgr}Estimates for the observables $s$ and $\theta_\infty$ defined in the text, for
the supermassive black hole candidate in M87, assuming a mass $M_{\text{M87}} = 6.5\times 10^{9} M_\odot$ and a distance $D_{OL} = 16.8\,\text{Mpc}$. For clarity we display the range $0\leq|\tilde Q_b|\leq0.5$. We remind the reader that $|\tilde{Q}_b|=\ell_{{b}}^2\gamma^2/R_{s}^{2}$.}
\centering
\begin{tabular}{ |c|c|c|c|c|c|c|c|} 
 \hline
 
  $|\tilde Q_b|$ & 0 & 0.1 & 0.2 & 0.3 & 0.4 & 0.5  \\ \hline
 $\theta_\infty\,(\mu\text{as})$ & 19.85 & 18.95 & 17.98 & 16.93 & 15.80 & 14.58  \\ 
 $s\,(\mu\text{as})$ & 0.0248 & 0.0289 & 0.0346 & 0.0422 & 0.0522 & 0.0654  \\

 \hline
\end{tabular}
\end{table}

\renewcommand{\arraystretch}{1.2}
\begin{table}
\caption{\label{table:m87}Estimates for the observables $s$ and $\theta_\infty$ defined in the text, for the supermassive black hole candidate \textit{Sgr A*}, assuming a mass $M_{Sgr} = 4.28\times 10^{6} M_\odot$ and a distance $D_{OL} = 8.32{\,\text{kpc}}$. For clarity, we display the range $0\leq|\tilde Q_b|\leq0.5$. We recall that $|\tilde{Q}_b|=\ell_{{b}}^2\gamma^2/R_s^2$.}
\centering
\begin{tabular}{ |c|c|c|c|c|c|c|c| } 
 \hline
 $|\tilde Q_b|$ & 0 & 0.1 & 0.2 & 0.3 & 0.4 & 0.5  \\ \hline
 $\theta_\infty\,(\mu\text{as})$ & 26.40 & 25.19 & 23.90 & 22.51 & 21.02 & 19.38  \\ 
 $s\,(\mu\text{as})$ & 0.0329 & 0.0384 & 0.046 & 0.0561 & 0.0695 & 0.087  \\
 \hline
\end{tabular}
\end{table}

\section{Concluding remarks}\label{Sec: Conclusions}
We have presented the first constraints on the free parameter $|\tilde{Q}_b|$ of the improved GUP scheme. Our results are derived from the analysis of light deflection in the strong-gravity regime, carried out using Bozza’s approximation. The validity of this method for static, spherically symmetric, and asymptotically flat spacetimes has been established in previous works (see, for instance,~\cite{Badia:2017art, Chagoya:2020bqz}). We found that, in the context of strong lensing, the parameter $|\tilde{Q}_b|$ is more relevant than $|\tilde{Q}_c|$, since the dominant correction to the photon sphere originates from $|\tilde{Q}_b|$ rather than from $|\tilde{Q}_c|$~\cite{PhysRevD.111.024048}. In particular, we found
that $|\tilde Q_b|\lesssim 0.3$ is required in order to 
have a shadow boundary radius that is compatible with EHT observations, both for Sgr A$^*$ and M87$^*$. Regarding the separation between images, in the range of $|\tilde Q_b|$ that we explored, it is twice as large for the GUP-improved black hole as for the Schwarzschild case, although still too small to be resolved by current telescopes. It is worth mentioning that even though Sgr A$^*$ and M87$^*$ are rotating black holes, their shadow radius is well approximated by that of a static, spherically symmetric spacetime. Therefore, we do not expect our constraints on $|\tilde Q_b|$ to receive large modifications due to rotational effects not considered in this work.

Consequently, due to the constraint on the value of $|\tilde{Q}_b|$, the effective horizon radius in Eq.~\eqref{eq_Eff_Rad_Horz} is also restricted according to the EHT observations. In particular, this horizon can take values within the range $0.84\,R_s \lesssim R_H \leq R_s$.

Another feature of the improved GUP black holes that was revealed by our analysis is that their photon capture radius is relatively smaller than that of a Schwarzschild black hole. It follows that if two static, spherically symmetric black holes share the same shadow boundary radius, one being Schwarzschild and the other arising from the improved GUP scheme, the latter would necessarily be more massive.

Our results also show that the deflection angle rapidly approaches that of a Schwarzschild black hole as one moves away from the photon sphere. This highlights the relevance of studying the strong deflection of light as a tool for constraining GUP deformation parameters.


\begin{acknowledgments}
 {\bf JC} is partially supported by DCF-320821.  {\bf IDS} and {\bf WY} are supported by the SECIHITI program ``Estancias Posdoctorales por M\' exico''. {\bf WY} thanks the SECIHTI grant CBF2023-2024-2923 ``Implications of the Generalized Uncertainty Principle (GUP) in Quantum Cosmology, Gravitation, and its Connection with Non-extensive Entropies''. {\bf BR} appreciates the support provided by SECIHITI through grants 932448.
\end{acknowledgments}

\section*{References}

\bibliographystyle{unsrt}
\bibliography{ref}

\begin{thebibliography}{10}

\bibitem{thiemann2008modern}
Thomas Thiemann.
\newblock {\em Modern canonical quantum general relativity}.
\newblock Cambridge University Press, 2008.

\bibitem{Abhay_Ashtekar_2003}
Abhay Ashtekar, Stephen Fairhurst, and Joshua~L Willis.
\newblock Quantum gravity, shadow states and quantum mechanics.
\newblock {\em Classical and Quantum Gravity}, 20(6):1031, feb 2003.

\bibitem{PhysRevD.76.044016}
Alejandro Corichi, Tatjana Vuka\v{s}inac, and Jos\'e. Zapata.
\newblock Polymer quantum mechanics and its continuum limit.
\newblock {\em Phys. Rev. D}, 76:044016, Aug 2007.

\bibitem{PhysRevD.95.065026}
Hugo~A. Morales-T\'ecotl, Saeed Rastgoo, and Juan~C. Ruelas.
\newblock Path integral polymer propagator of relativistic and nonrelativistic particles.
\newblock {\em Phys. Rev. D}, 95:065026, Mar 2017.

\bibitem{PhysRevD.92.104029}
Hugo~A. Morales-T\'ecotl, Daniel~H. Orozco-Borunda, and Saeed Rastgoo.
\newblock Polymer quantization and the saddle point approximation of partition functions.
\newblock {\em Phys. Rev. D}, 92:104029, Nov 2015.

\bibitem{FLORESGONZALEZ2013394}
Ernesto Flores-Gonz\'alez, Hugo~A. Morales-T\'ecotl, and Juan~D. Reyes.
\newblock Propagators in polymer quantum mechanics.
\newblock {\em Annals of Physics}, 336:394--412, 2013.

\bibitem{PhysRevLett.110.211301}
Rodolfo Gambini and Jorge Pullin.
\newblock Loop quantization of the schwarzschild black hole.
\newblock {\em Phys. Rev. Lett.}, 110:211301, May 2013.

\bibitem{Corichi_2016}
Alejandro Corichi and Parampreet Singh.
\newblock Loop quantization of the schwarzschild interior revisited.
\newblock {\em Classical and Quantum Gravity}, 33(5):055006, feb 2016.

\bibitem{PhysRevD.78.064040}
Dah-Wei Chiou.
\newblock Phenomenological loop quantum geometry of the schwarzschild black hole.
\newblock {\em Phys. Rev. D}, 78:064040, Sep 2008.

\bibitem{MORALESTECOTL2021168401}
Hugo~A. Morales-T\'ecotl, Saeed Rastgoo, and Juan~C. Ruelas.
\newblock Effective dynamics of the schwarzschild black hole interior with inverse triad corrections.
\newblock {\em Annals of Physics}, 426:168401, 2021.

\bibitem{PhysRevD.76.104030}
Christian~G. B\"ohmer and Kevin Vandersloot.
\newblock Loop quantum dynamics of the schwarzschild interior.
\newblock {\em Phys. Rev. D}, 76:104030, Nov 2007.

\bibitem{https://doi.org/10.1155/2008/459290}
Leonardo Modesto.
\newblock Black hole interior from loop quantum gravity.
\newblock {\em Advances in High Energy Physics}, 2008(1):459290, 2008.

\bibitem{PhysRevD.103.084038}
Keagan Blanchette, Saurya Das, Samantha Hergott, and Saeed Rastgoo.
\newblock Black hole singularity resolution via the modified raychaudhuri equation in loop quantum gravity.
\newblock {\em Phys. Rev. D}, 103:084038, Apr 2021.

\bibitem{Sobrinho_2023}
F~C Sobrinho, H~A Borges, I~P~R Baranov, and S~Carneiro.
\newblock On the horizon area of effective loop quantum black holes.
\newblock {\em Classical and Quantum Gravity}, 40(14):145003, jun 2023.

\bibitem{Adeifeoba_2018}
Ad\'em\'ol\'a Ad\'eif\'eoba, Astrid Eichhorn, and Alessia~Benedetta Platania.
\newblock Towards conditions for black-hole singularity-resolution in asymptotically safe quantum gravity.
\newblock {\em Classical and Quantum Gravity}, 35(22):225007, oct 2018.

\bibitem{olmo2016classical}
Gonzalo~J Olmo, D~Rubiera-Garcia, and A~Sanchez-Puente.
\newblock Classical resolution of black hole singularities via wormholes.
\newblock {\em The European Physical Journal C}, 76:1--6, 2016.

\bibitem{Husain_2005}
Viqar Husain and Oliver Winkler.
\newblock Quantum resolution of black hole singularities.
\newblock {\em Classical and Quantum Gravity}, 22(21):L127, oct 2005.

\bibitem{PhysRevD.73.104009}
Daniel Cartin and Gaurav Khanna.
\newblock Wave functions for the schwarzschild black hole interior.
\newblock {\em Phys. Rev. D}, 73:104009, May 2006.

\bibitem{jalalzadeh2012quantization}
Shahram Jalalzadeh and Babak Vakili.
\newblock Quantization of the interior schwarzschild black hole.
\newblock {\em International Journal of Theoretical Physics}, 51:263--275, 2012.

\bibitem{PhysRevD.57.2349}
Thorsten Brotz.
\newblock Quantization of black holes in the wheeler-dewitt approach.
\newblock {\em Phys. Rev. D}, 57:2349--2362, Feb 1998.

\bibitem{Almeida_2023}
Carla~R Almeida and Denis~C Rodrigues.
\newblock Quantization of a black-hole gravity: geometrodynamics and the quantum.
\newblock {\em Classical and Quantum Gravity}, 40(3):035004, jan 2023.

\bibitem{PhysRevD.60.024009}
Cenalo Vaz and Louis Witten.
\newblock Mass quantization of the schwarzschild black hole.
\newblock {\em Phys. Rev. D}, 60:024009, Jun 1999.

\bibitem{PhysRevLett.98.181301}
Alejandro Corichi, Jacobo D\'{\i}az-Polo, and Enrique Fern\'andez-Borja.
\newblock Black hole entropy quantization.
\newblock {\em Phys. Rev. Lett.}, 98:181301, May 2007.

\bibitem{PhysRevD.81.023528}
A.~Bina, S.~Jalalzadeh, and A.~Moslehi.
\newblock Quantum black hole in the generalized uncertainty principle framework.
\newblock {\em Phys. Rev. D}, 81:023528, Jan 2010.

\bibitem{PhysRevD.52.1108}
Achim Kempf, Gianpiero Mangano, and Robert~B. Mann.
\newblock Hilbert space representation of the minimal length uncertainty relation.
\newblock {\em Phys. Rev. D}, 52:1108--1118, Jul 1995.

\bibitem{PhysRevD.85.024016}
Pouria Pedram.
\newblock New approach to nonperturbative quantum mechanics with minimal length uncertainty.
\newblock {\em Phys. Rev. D}, 85:024016, Jan 2012.

\bibitem{PhysRevD.107.126009}
Pasquale Bosso, Luciano Petruzziello, and Fabian Wagner.
\newblock Minimal length: A cut-off in disguise?
\newblock {\em Phys. Rev. D}, 107:126009, Jun 2023.

\bibitem{BOSSO2018498}
Pasquale Bosso, Saurya Das, and Robert~B. Mann.
\newblock Potential tests of the generalized uncertainty principle in the advanced ligo experiment.
\newblock {\em Physics Letters B}, 785:498--505, 2018.

\bibitem{BIZET2023137636}
Nana~Cabo Bizet, Octavio Obreg\'{o}n, and Wilfredo Yupanqui.
\newblock Modified entropies as the origin of generalized uncertainty principles.
\newblock {\em Physics Letters B}, 836:137636, 2023.

\bibitem{G_Veneziano_1986}
G.~Veneziano.
\newblock A stringy nature needs just two constants.
\newblock {\em Europhysics Letters}, 2(3):199, aug 1986.

\bibitem{MAGGIORE199365}
Michele Maggiore.
\newblock A generalized uncertainty principle in quantum gravity.
\newblock {\em Physics Letters B}, 304(1):65--69, 1993.

\bibitem{SCARDIGLI199939}
Fabio Scardigli.
\newblock Generalized uncertainty principle in quantum gravity from micro-black hole gedanken experiment.
\newblock {\em Physics Letters B}, 452(1):39--44, 1999.

\bibitem{Bosso_2023}
Pasquale Bosso, Giuseppe~Gaetano Luciano, Luciano Petruzziello, and Fabian Wagner.
\newblock 30 years in: Quo vadis generalized uncertainty principle?
\newblock {\em Classical and Quantum Gravity}, 40(19):195014, sep 2023.

\bibitem{Bosso_2021}
Pasquale Bosso, Octavio Obregón, Saeed Rastgoo, and Wilfredo Yupanqui.
\newblock Deformed algebra and the effective dynamics of the interior of black holes.
\newblock {\em Classical and Quantum Gravity}, 38(14):145006, jun 2021.

\bibitem{MELCHOR2024116584}
Brayan Melchor, Rolando Perca, and Wilfredo Yupanqui.
\newblock Semiclassical resolution of the black hole singularity inspired in the minimal uncertainty approach.
\newblock {\em Nuclear Physics B}, 1004:116584, 2024.

\bibitem{PhysRevD.65.125028}
Lay~Nam Chang, Djordje Minic, Naotoshi Okamura, and Tatsu Takeuchi.
\newblock Effect of the minimal length uncertainty relation on the density of states and the cosmological constant problem.
\newblock {\em Phys. Rev. D}, 65:125028, Jun 2002.

\bibitem{PhysRevD.66.026003}
S\'andor Benczik, Lay~Nam Chang, Djordje Minic, Naotoshi Okamura, Saiffudin Rayyan, and Tatsu Takeuchi.
\newblock Short distance versus long distance physics: The classical limit of the minimal length uncertainty relation.
\newblock {\em Phys. Rev. D}, 66:026003, Jun 2002.

\bibitem{CASADIO2020135558}
Roberto Casadio and Fabio Scardigli.
\newblock Generalized uncertainty principle, classical mechanics, and general relativity.
\newblock {\em Physics Letters B}, 807:135558, 2020.

\bibitem{PhysRevD.111.024048}
Federica Fragomeno, Douglas~M. Gingrich, Samantha Hergott, Saeed Rastgoo, and Evan Vienneau.
\newblock A generalized uncertainty-inspired quantum black hole.
\newblock {\em Phys. Rev. D}, 111:024048, Jan 2025.

\bibitem{Bosso}
Pasquale Bosso, Octavio Obreg\'on, Saeed Rastgoo, and Wilfredo Yupanqui.
\newblock Deformed algebra and the effective dynamics of the interior of black holes.
\newblock {\em Class. Quantum Grav.}, 38(14):145006, 2021.
\newblock 19pp.

\bibitem{doi:10.1098/rspa.1959.0015}
Charles~Galton Darwin.
\newblock The gravity field of a particle.
\newblock {\em Proceedings of the Royal Society of London. Series A. Mathematical and Physical Sciences}, 249(1257):180--194, 1959.

\bibitem{Akiyama:2019cqa}
Kazunori Akiyama et~al.
\newblock {First M87 Event Horizon Telescope Results. I. The Shadow of the Supermassive Black Hole}.
\newblock {\em Astrophys. J.}, 875(1):L1, 2019.

\bibitem{2022ApJ...930L..12E}
{Event Horizon Telescope Collaboration}.
\newblock {First Sagittarius A* Event Horizon Telescope Results. I. The Shadow of the Supermassive Black Hole in the Center of the Milky Way}.
\newblock {\em "ApJL"}, 930(2):L12, May 2022.

\bibitem{Bozza_2002}
V.~Bozza.
\newblock {Gravitational lensing in the strong field limit}.
\newblock {\em Phys. Rev. D}, 66:103001, 2002.

\bibitem{Virbhadra:1998dy}
K.S. Virbhadra, D.~Narasimha, and S.M. Chitre.
\newblock {Role of the scalar field in gravitational lensing}.
\newblock {\em Astron. Astrophys.}, 337:1--8, 1998.

\bibitem{Wei_2015}
Shao-Wen Wei, Ke~Yang, and Yu-Xiao Liu.
\newblock {Black hole solution and strong gravitational lensing in Eddington-inspired Born\textendash{}Infeld gravity}.
\newblock {\em Eur. Phys. J. C}, 75:253, 2015.
\newblock [Erratum: Eur.Phys.J.C 75, 331 (2015)].

\bibitem{Zhao:2016kft}
Shan-Shan Zhao and Yi~Xie.
\newblock {Strong field gravitational lensing by a charged Galileon black hole}.
\newblock {\em JCAP}, 07:007, 2016.

\bibitem{Badia:2017art}
Javier Bad\'ia and Ernesto~F. Eiroa.
\newblock {Gravitational lensing by a Horndeski black hole}.
\newblock {\em Eur. Phys. J. C}, 77(11):779, 2017.

\bibitem{Izmailov:2019uhy}
R.N. Izmailov, R.Kh. Karimov, E.R. Zhdanov, and K.K. Nandi.
\newblock {Modified gravity black hole lensing observables in weak and strong field of gravity}.
\newblock {\em Mon. Not. Roy. Astron. Soc.}, 483(3):3754--3761, 2019.

\bibitem{Chagoya:2020bqz}
Javier Chagoya, C.~Ortiz, Benito Rodr\'\i{}guez, and Armando~A. Roque.
\newblock {Strong gravitational lensing by DHOST black holes}.
\newblock {\em Class. Quant. Grav.}, 38(7):075026, 2021.

\bibitem{scardigli2015gravitational}
Fabio Scardigli and Roberto Casadio.
\newblock Gravitational tests of the generalized uncertainty principle.
\newblock {\em The European Physical Journal C}, 75(9):1--12, 2015.

\bibitem{OKCU2021115324}
Özgür Ökcü and Ekrem Aydiner.
\newblock Observational tests of the generalized uncertainty principle: Shapiro time delay, gravitational redshift, and geodetic precession.
\newblock {\em Nuclear Physics B}, 964:115324, 2021.

\bibitem{TURAKHONOV2024101716}
Ziyodulla Turakhonov, Husanboy Hoshimov, Farruh Atamurotov, Sushant~G. Ghosh, and Ahmadjon Abdujabbarov.
\newblock Observational signatures of strong gravitational lensing in gup-modified schwarzschild black holes.
\newblock {\em Physics of the Dark Universe}, 46:101716, 2024.

\bibitem{PhysRevD.90.024023}
Souvik Pramanik.
\newblock Implication of the geodesic equation in the generalized uncertainty principle framework.
\newblock {\em Phys. Rev. D}, 90:024023, Jul 2014.

\bibitem{Ashtekar_2006}
Abhay Ashtekar and Martin Bojowald.
\newblock Quantum geometry and the schwarzschild singularity.
\newblock {\em Classical and Quantum Gravity}, 23(2):391, dec 2005.

\bibitem{Thiemann_2007}
Thomas Thiemann.
\newblock {\em Modern Canonical Quantum General Relativity}.
\newblock Cambridge Monographs on Mathematical Physics. Cambridge University Press, 2007.

\bibitem{Bosso_2024}
Pasquale Bosso, Octavio Obregón, Saeed Rastgoo, and Wilfredo Yupanqui.
\newblock Black hole interior quantization: a minimal uncertainty approach.
\newblock {\em Classical and Quantum Gravity}, 41(13):135011, jun 2024.

\bibitem{PhysRevD.79.125007}
M.~Zarei and B.~Mirza.
\newblock Minimal uncertainty in momentum: The effects of ir gravity on quantum mechanics.
\newblock {\em Phys. Rev. D}, 79:125007, Jun 2009.

\bibitem{MUREIKA201988}
J.R. Mureika.
\newblock Extended uncertainty principle black holes.
\newblock {\em Physics Letters B}, 789:88--92, 2019.

\bibitem{doi:10.1142/S0217732319502043}
Won~Sang Chung, Hassan Hassanabadi, and Nasrin Farahani.
\newblock Klein–gordon oscillator in the presence of the minimal momentum.
\newblock {\em Modern Physics Letters A}, 34(25):1950204, 2019.

\bibitem{Bosso_2020}
Pasquale Bosso and Octavio Obregón.
\newblock Minimal length effects on quantum cosmology and quantum black hole models.
\newblock {\em Classical and Quantum Gravity}, 37(4):045003, jan 2020.

\bibitem{PhysRevD.74.084003}
Abhay Ashtekar, Tomasz Pawlowski, and Parampreet Singh.
\newblock Quantum nature of the big bang: Improved dynamics.
\newblock {\em Phys. Rev. D}, 74:084003, Oct 2006.

\bibitem{chiou2013loopquantizationsphericallysymmetric}
Dah-Wei Chiou, Wei-Tou Ni, and Alf Tang.
\newblock Loop quantization of spherically symmetric midisuperspaces and loop quantum geometry of the maximally extended schwarzschild spacetime, 2013.

\bibitem{Virbhadra:2002ju}
K.~S. Virbhadra and G.~F.~R. Ellis.
\newblock {Gravitational lensing by naked singularities}.
\newblock {\em Phys. Rev. D}, 65:103004, 2002.

\bibitem{Weinberg:100595}
Steven Weinberg.
\newblock {\em {Gravitation and Cosmology: Principles and Applications of the General Theory of Relativity}}.
\newblock Wiley, New York, NY, 1972.

\bibitem{Akiyama:2019eap}
Kazunori Akiyama et~al.
\newblock {First M87 Event Horizon Telescope Results. VI. The Shadow and Mass of the Central Black Hole}.
\newblock {\em Astrophys. J. Lett.}, 875(1):L6, 2019.

\bibitem{Gillessen_2017}
S.~Gillessen, P.~M. Plewa, F.~Eisenhauer, R.~Sari, I.~Waisberg, M.~Habibi, O.~Pfuhl, E.~George, J.~Dexter, S.~von Fellenberg, and et~al.
\newblock An update on monitoring stellar orbits in the galactic center.
\newblock {\em Astrophys. J.}, 837(1):30, 2017.

\end{thebibliography}
        
\end{document}